\newcommand{\be}{\begin{eqnarray}}
\newcommand{\ee}{\end{eqnarray}}

\def\bd{\begin{displaymath}}
\def\ed{\end{displaymath}}

\def\ba#1{\begin{array}{#1}}
\def\ea{\end{array}}

\newfont{\Bbb}{msbm10 scaled 1200}
\documentclass[12pt]{article}
\usepackage{feynmp}
\usepackage{amsmath}
\newcommand{\mcom}{,}
\usepackage{fleqn}
\usepackage{enumerate}
\title{On the nullification of threshold amplitudes}

\author{
Joanna Gonera\thanks{supported by the Lodz University Grant N$^o$442} \\
{Department of Theoretical Physics II, University of \L\'od\'z,}\\
{ul. Pomorska 149/153, 90236 \L\'od\'z, Poland.}
}
\date{}
\begin{document}
\maketitle{}

\begin{abstract}
The nullification of threshold amplitudes is considered within the conventional framework
of quantum field theory. The relevant Ward identities for the reduced theory are derived both
on path-integral and diagrammatic levels. They are then used to prove the vanishing of tree-graph 
threshold amplitudes.
\end{abstract}
\section{Introduction}
Threshold amplitudes have attracted much attention in the last decade ( for a review see \cite{b1}).
In particular, the tree-graph approximation has been analysed in great detail \cite{b2}. One
of the most interesting phenomena discovered here is the so-called nullification of amplitudes;
it appeared that, in some theories, a number of threshold amplitudes vanish. For example, in 
unbroken $\Phi^4$-theory all threshold amplitudes $2 \rightarrow n$\ vanish except $n=2$\ and
$n=4$\ \cite{b3}; in the case of spontaneusly broken $\Phi \rightarrow -\Phi$\ symmetry the
only nonvanishing amplitude is the $2\rightarrow 2$\ one \cite{b4}. Other examples concern theories 
with more bosonic fields where the nullification phenomenon occurs provided certain relations
between the parameters of the theory hold \cite{b5}. \\
Particularly interesting nullification phenomena occur in bosonic theory with softly broken
$O(2)$\ symmetry with two real fields in basic representation of $O(2)$. The relevant 
lagrangian reads
\be
L=\frac{1}{2}((\partial_{\mu}\Phi_1)^2+(\partial_{\mu}\Phi_2)^2-m_1^2\Phi_1^2-m_2^2\Phi_2^2)
-\frac{\lambda}{4!}(\Phi^2_1+\Phi_2^2)^2, \label{w1}
\ee
with $ m_1 \neq m_2$. It has been found that the tree amplitudes $n_1\Phi_1\rightarrow n_2\Phi_2$,
with all particles (initial as well as final) on the threshold, vanish \cite{b6}. This result, 
obtained in Ref.\cite{b6} by explicit calculations, admits more general and interesting explanation
\cite{b7}. The generating functional for tree amplitudes is obtained  by solving the 
classical field equations. For threshold amplitudes they reduce to one-dimensional equations
for dynamical system of two degrees of freedom. One can show that the nonvanishing amplitudes 
can occur only if the resonances appear when the equations of motion are solved perturbatively.
Now, the reduced system posses a special symmetry which exludes such possibility.
This relation between symmetries and threshold nullification phenomena is more general and
can be described  within the framework of modern theory of integrable systems \cite{b8}.\\
On the other hand the nullification is ultimately a result of subtle cancellations between 
contributions coming from different tree graphs. One can suspect that these cancellations result
from Ward identities related to the symmetry of reduced system. This more traditional point of view
is the subject of the present paper. We derive the Ward identities for the reduced system, both
by path-integral and diagrammatical methods. Then the coincidence of threshold tree amplitudes
with those of reduced theory is used to show that the tree-level contributions do cancel.
\section{Ward identities}
We start with the hamiltonian of the reduced system corresponding to translational-invariant version
of eq.(\ref{w1})
\be
H=\frac{1}{2}(\Pi_1^2+m_1^2\varphi_1^2)+\frac{1}{2}(\Pi_2^2+m_2^2\varphi_2^2)+
\frac{\lambda}{4!}(\varphi^2_1+\varphi_2^2)^2,\label{w2}
\ee
where $\Pi_i \equiv \dot{\varphi_i},\;i=1,2$. The system is integrable \cite{b9} \cite{b10},
 the two independent commuting 
integrals being
\be
F_i=\frac{\lambda J^2}{4!\Delta m_i^2}+\frac{1}{2}(\Pi^2_i+m_i^2\varphi_i^2)+
\frac{\lambda}{4!}\varphi_i^2({\vec{\varphi}}^2);\label{w3}
\ee
here ${\vec{\varphi}}^2\equiv \varphi_1^2+\varphi_2^2,\; \Delta m_i^2\equiv m_j^2-m_i^2, \;
i\neq j$\ and 
\be
J \equiv \sum_{i,j}\varepsilon_{ij}\varphi_i\Pi_j=\varphi_1\Pi_2-\varphi_2\Pi_1 \nonumber
\ee
is two-dimensional angular momentum. The existence of two integrals quadratic in momenta is 
implied by the separability of the potential (in eliptic coordinates \cite{b10}). \\
For any pair $\alpha_1,\alpha_2\in R$\ one can define the generator of symmetry transformations
\be
F_{(\alpha )}\equiv \alpha_1 F_1+\alpha_2F_2;\label{w4}
\ee
in particular $F_{(1,1)}=H$. Let us define for further use
\be
&&\alpha \equiv \alpha_1+\alpha_2 \nonumber \\
&&\beta \equiv \sum_{i=1}^2\frac{\alpha_i}{\Delta m_i^2}=\frac{\alpha_1-\alpha_2}
{m_2^2-m_1^2}\label{w5}
\ee
The relevant symmetry transformations read (in infinitesimal form)  
\be
&&\varphi_i\rightarrow \varphi_i' =\varphi_i +\varepsilon \{\varphi_i,F_{(\alpha)}\}=\varphi_i+
\varepsilon (\alpha_i\Pi_i-\frac{2\lambda \beta}{4!}J\sum_{j=1}^2\varepsilon_{ij}\varphi_j)\equiv\nonumber \\
&&\equiv \varphi_i+\varepsilon Q_i^{(\alpha)}\label{w6} \\
&&\Pi_i\rightarrow \Pi_i'=\Pi_i+\varepsilon \{\Pi_i,F_{(\alpha)}\}=\Pi_i+\varepsilon 
(-\alpha_im_i^2\varphi_i-\frac{2\lambda}{4!}\alpha_i\varphi_i{\vec{\phi}}^2 +\nonumber \\
&&-\frac{2\lambda}{4!}\varphi_i(\sum_{j=1}^2\alpha_j\varphi_j^2)-\frac{2\lambda \beta}{4!}
J\sum_{j=1}^2\varepsilon_{ij}\Pi_j)\equiv \nonumber \\
&&\equiv \Pi_i+\varepsilon P_i^{(\alpha)} \label{w7}
\ee
Let us remind the Noether theorem in the hamiltonian framework. The canonical transformation
$\varphi_i \rightarrow \varphi_i',\; \Pi_i\rightarrow \Pi_i'$\ is a symmetry transormation if the 
following condition holds
\be
\sum_i\Pi_i' \dot{\varphi_i'}-H(\varphi ',\Pi ')=\sum_i\Pi_i\dot{\varphi_i}-H(\varphi ,\Pi )+
\dot{\tilde{\Psi }}(\varphi ,\Pi ), \label{w8}
\ee
where $\tilde {\Psi} (\varphi ,\Pi )$\ is some function (in general it may depend also on time). Taking 
an infinitesimal form of symmetry transformations
\be
&&\varphi_i'=\varphi_i+\varepsilon Q_i \nonumber \\
&&\Pi_i'=\Pi_i+\varepsilon P_i \label{w9} \\
&&\tilde {\Psi} (\varphi ,\Pi )=\varepsilon \Psi (\varphi ,\Pi ) \nonumber
\ee
one obtains from eq.(\ref{w8}) 
\be
\sum_i(\dot{\varphi_i}-\frac{\partial H}{\partial \Pi_i})P_i-\sum_i(\dot{\Pi_i}+\frac{\partial H}{
\partial \varphi_i})Q_i+\frac{d}{dt}(\sum_i\Pi_iQ_i-\Psi )=0 \label{w10}.
\ee
This is the hamiltonian form of Noether theorem. It is easy to check that the conserved 
quantity generates the symmetry transformations. Indeed, eq.(\ref{w8}) implies
\be
\sum_i\Pi_i'd\varphi_i'-\sum_i\Pi_id\varphi_i=d\Psi (\varphi ,\Pi ) \label{w11}
\ee
or, infinitesimally, 
\be
\sum_i(P_id\varphi_i+\Pi_idQ_i)=d\Psi \label{w12}
\ee
so that
\be
&&P_i+\sum_j\Pi_j\frac{\partial Q_j}{\partial \varphi_i}=\frac{\partial \Psi}{\partial \varphi_i} \label{w13}\\
&&\sum_j\Pi_j\frac{\partial Q_j}{\partial \Pi_i}=\frac{\partial \Psi}{\partial \Pi_i}\;. \nonumber
\ee
Eqs.(\ref{w13}) can be written as 
\be
&&\{ \varphi_i,\;\sum_j\Pi_jQ_j-\Psi \} =Q_i \nonumber \\
&&\{ \Pi_i, \; \sum_j \Pi_jQ_j-\Psi \} =P_i \nonumber
\ee
which imply that $F\equiv \sum_j\Pi_jQ_j-\Psi $\ is a conserved  generator of symmetry transformations.\\
Now, we can prove the relevant Ward identities. We start with path-integral representation of the
generating functional
\be
Z[\vec{J},\vec{K}]=\int D\varphi 'D\Pi 'e^{i \int (\sum_i\Pi_i'\dot{\varphi_i}'-H(\varphi ',
\Pi '))dt+i \int \sum_i(J_i\varphi_i' +K_i\Pi_i')dt}\label{w14}
\ee
Let us make a canonical transformation 
\be
&&\varphi_i'=\varphi_i+\varepsilon Q_i \nonumber \\
&&\Pi_i'=\Pi_i+\varepsilon P_i \label{w15}
\ee
with time-dependent parameter $\varepsilon =\varepsilon (t)$. Canonicity implies
formal measure invariance, $D\varphi ' D\Pi '=D\varphi D\Pi $. Moreover, 
assuming that for a constant $\varepsilon$ (\ref{w15}) is a symmetry transformation, one gets
\be
\sum_i\Pi_i'\dot{\varphi_i}'-H(\varphi ',\Pi ')=\sum_i\Pi_i\dot{\varphi_i}-H(\varphi ,\Pi )+
\dot{\varepsilon}\sum_i\Pi_iQ_i+\varepsilon \dot{\Psi}\label{w16}
\ee
Therefore
\be
&&Z[\vec{J},\vec{K}]=\int D\varphi D\Pi e^{i\int (\sum_i\Pi_i\dot{\varphi_i}-H)dt+
i\int \sum_i(J_i\varphi_i+K_i\Pi_i)dt }\cdot \nonumber \\
&&\cdot e^{i\int (\sum_i(\dot{\varepsilon} \Pi_iQ_i+\varepsilon J_iQ_i+\varepsilon K_iP_i)
+\varepsilon \dot{\Psi})dt} \label{w17}
\ee
or, to the first order,
\be
&&\int D\varphi D\Pi(\int (\sum_i(\dot{\varepsilon} \Pi_iQ_i+\varepsilon J_iQ_i+\varepsilon K_iP_i)
+\varepsilon \dot{\Psi})dt)\cdot \nonumber \\
&&\cdot e^{i\int (\sum_i\Pi_i\dot{\varphi_i}-H)dt+i\int \sum_i(J_i
\varphi_i+K_i\Pi_i)dt}
=0 \label{w18}
\ee
Assuming that $\varepsilon (t)$\ vanishes outside a finite interval and integrating by parts we arrive finally at
\begin{eqnarray}
\lefteqn{\int D\varphi D\Pi(\sum_i (J_iQ_i+K_iP_i)(t)+{}}&{}&{}\nonumber\\
&-&\dot{F}(t))e^{i\int (\sum_i\Pi_i\dot{\varphi_i}-H)dt
+i\int \sum_i (J_i\varphi_i +K_i\Pi_i)dt }=0 \label{w19}
\end{eqnarray}
Differentiating $n$\ times with respect to $J_i$\ and putting $\vec{J}=0,\; \vec{K}=0$\ we get
\be
&&\frac{d}{dt}<T(F(t))\varphi_{i_1}(t_1)...\varphi_{i_n}(t_n))>=\frac{1}{i}\sum_{k=1}^n
\delta(t-t_k) \cdot \nonumber \\
&&<T(Q_{i_k}(t)\varphi_{i_1}(t_1)...\overset{\textstyle\vee}{\varphi_{i_k}(t_k)}...\varphi_{i_n}(t_n))>\label{w20}
\ee
These are Ward identities following from the symmetry. More general identities can be obtained by 
differentiating with respect to $\vec{J}$\ and $\vec{K}$\ but we shall not need them here. 
The Ward identities (\ref{w20}) can be also obtained from canonical commutation rules and 
Heisenberg equations of motion.

\section{Diagrammatical proof}
The derivation given above is slightly formal. It is desirable to give a diagrammatical proof which allows for a better 
insight into the cancellation mechanism which makes the amplitudes vanishing.

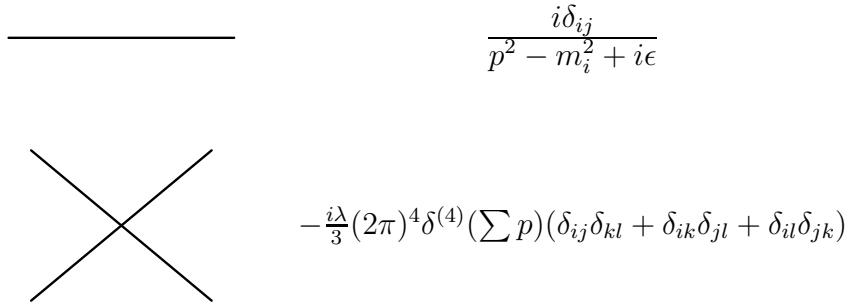
\begin{figure}
\[
\begin{array}{cc}
\parbox{4cm}{
   \fmfframe(5,5)(5,5){
     \begin{fmfgraph*}(30,10)
        \fmfleft{i1}\fmfright{o1}
        \fmf{plain}{i1,o1}
        \fmflabel{$i$}{i1}
        \fmflabel{$j$}{o1}
     \end{fmfgraph*}
   }
} &
\frac{\textstyle i\delta_{ij}}{\textstyle p^2-m_i^2+i\epsilon}
\\
\parbox{4cm}{
\fmfframe(5,5)(5,5){
\begin{fmfgraph*}(30,20)
\fmfleft{i1,i2} \fmfright{o1,o2}
\fmf{plain}{i1,o2} \fmf{plain}{i2,o1}
\fmflabel{$l$}{i1}
\fmflabel{$i$}{i2}
\fmflabel{$k$}{o1}
\fmflabel{$j$}{o2}
\end{fmfgraph*}}
} &
-\frac{i\lambda}{3}(2\pi)^4\delta^{(4)}(\sum p)(\delta_{ij}\delta_{kl}+\delta_{ik}\delta_{jl}+
\delta_{il}\delta_{jk})
\end{array}
\]
\caption{Feynman rules for $\phi^4$ theory}
\label{rul0}
\end{figure}

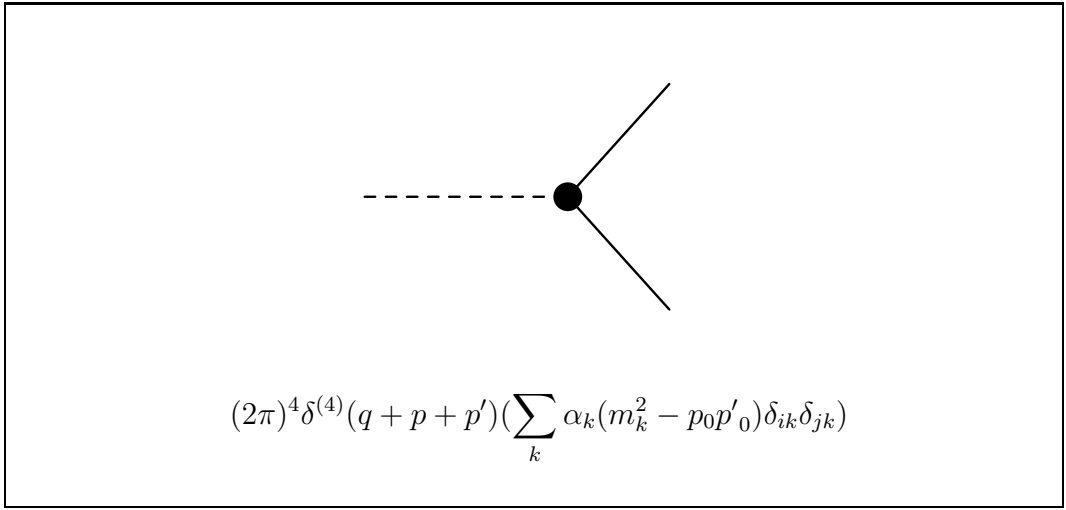
\begin{figure} 
\begin{center}
\fbox{
\parbox{\textwidth}{
\begin{center}
\fmfframe(5,5)(5,5){
\begin{fmfgraph*}(45,30)
  \fmfleft{i1} \fmfright{o1,o2}
  \fmf{dashes,label=$q$}{i1,v1}
  \fmf{plain}{v1,o1}
  \fmf{plain}{v1,o2}
  \fmfv{decor.shape=circ,decor.filled=full,decor.size=5thick}{v1}
  \fmflabel{$p',j$}{o2}
  \fmflabel{$p,i$}{o1}
\end{fmfgraph*}
}\\
{\ }\\
\({\displaystyle
(2\pi)^4\delta^{(4)}(q+p+p')(\sum_k\alpha_k(m_k^2-p_0{p'}_0)\delta_{ik}\delta_{jk})}
\)
\end{center}
}}
\end{center}
\caption{Rule for $F^{(0)}_{(\alpha)}$}
\label{rul1}
\end{figure}

\begin{figure}
\begin{center}
\fbox{\parbox{\textwidth}{
\begin{center}
\fmfframe(5,5)(5,5){
\begin{fmfgraph*}(45,30)
  \fmfleft{i1} \fmfright{o1,o2,o3,o4}
  \fmf{dashes,label=$q$}{i1,v1}
  \fmf{plain}{v1,o1}
  \fmf{plain}{v1,o2}
  \fmf{plain}{v1,o3}
  \fmf{plain}{v1,o4}
  \fmfv{decor.shape=circ,decor.filled=full,decor.size=5thick}{v1}
  \fmflabel{$p_1,i_1$}{o4}
  \fmflabel{$p_2,i_2$}{o3}
  \fmflabel{$p_3,i_3$}{o2}
  \fmflabel{$p_4,i_4$}{o1}
\end{fmfgraph*}
}
\\{\ }\\
\(
\begin{array}{l}
{\displaystyle (2\pi)^4\delta^{(4)}(q+\sum p)(-\frac{2\beta\lambda}{4!}(\delta_{i_1i_2}\delta_{i_3i_4}
((p_{10}-p_{30})(p_{20}-p_{40})+{}}
\\{}{\displaystyle +(p_{20}-p_{30})(p_{10}-p_{40}))
+\frac{\lambda}{3!}\sum_k\alpha_k(\delta_{ki_1}\delta_{ki_2}\delta_{i_3i_4}+
\delta_{ki_3}\delta_{ki_4}\delta_{i_1i_2})+{}}\\ {\displaystyle {}+\mbox{\rm perm.})}
\end{array}
\)
\end{center}}}
\end{center}
\caption{Rule for $F^{(1)}_{(\alpha)}$}
\label{rul2}
\end{figure}
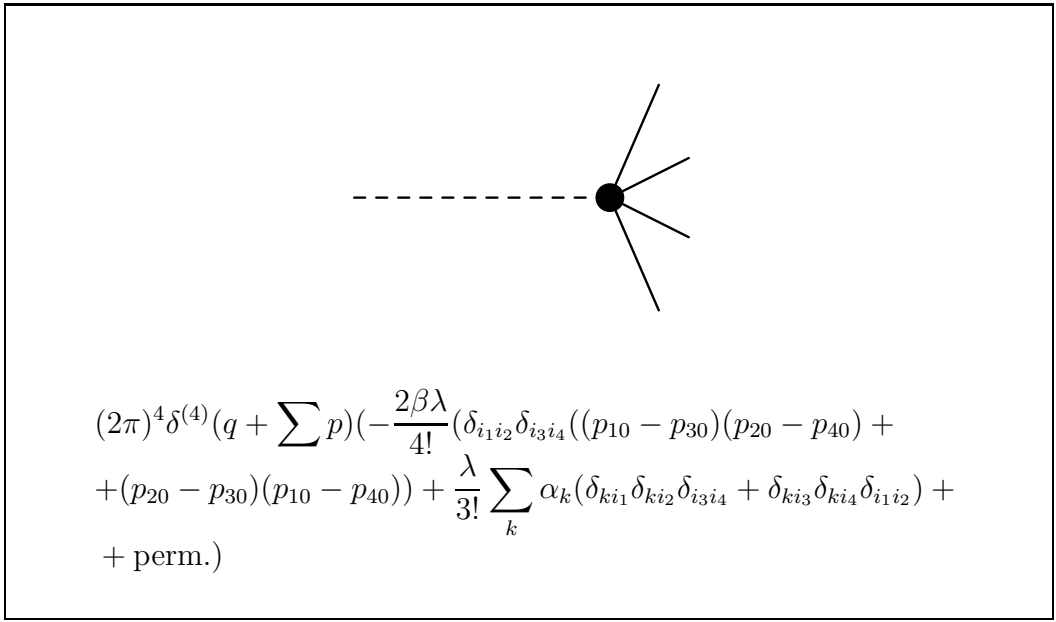

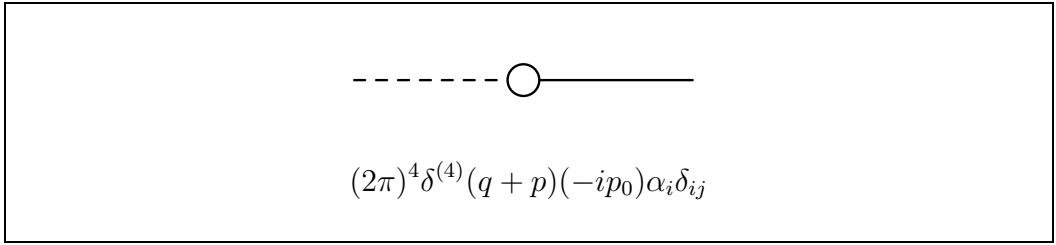
\begin{figure}
\begin{center}
\fbox{\parbox{\textwidth}{
\begin{center}
\fmfframe(5,2)(5,2){
\begin{fmfgraph*}(45,5)
  \fmfleft{i1} \fmfright{o1}
  \fmf{dashes}{i1,v1}
  \fmf{plain}{v1,o1}
  \fmflabel{$q,i$}{i1}
  \fmflabel{$p,j$}{o1}
  \fmfv{decor.shape=circle,decor.filled=empty,decor.size=6thick}{v1}
\end{fmfgraph*}
}\\{\ }\\
\({\displaystyle
(2\pi)^4\delta^{(4)}(q+p)(-ip_0)\alpha_i\delta_{ij}}
\)
\end{center}}}
\end{center}
\caption{Rule for $Q^{(0)}_{(\alpha)i}$}
\label{rul3}
\end{figure}

\begin{figure}
\begin{center}
\fbox{\parbox{\textwidth}{
\begin{center}
\fmfframe(5,5)(5,5){
\begin{fmfgraph*}(45,30)
  \fmfleft{i1} \fmfright{o1,o2,o3}
  \fmf{dashes}{i1,v1}
  \fmf{plain}{v1,o1}
  \fmf{plain}{v1,o2}
  \fmf{plain}{v1,o3}
  \fmflabel{$q,i$}{i1}
  \fmflabel{$p_1,i_3$}{o3}
  \fmflabel{$p_2,i_2$}{o2}
  \fmflabel{$p_3,i_3$}{o1}
  \fmfv{decor.shape=circle,decor.filled=empty,decor.size=6thick}{v1}
\end{fmfgraph*}
}\\{\ }\\
\({\displaystyle
(2\pi)^4\delta^{(4)}(q+\sum p)(\frac{2i\beta\lambda}{4!}(p_{20}+p_{30})\delta_{ii_1}\delta_{i_2i_3}
-\frac{i\lambda\beta}{3!}p_{10}\delta_{ii_1}\delta_{i_2i_3}+\mbox{\rm perm})}
\)
\end{center}}}
\end{center}
\caption{Rule for $Q^{(1)}_{(\alpha)i}$}
\label{rul4}
\end{figure}
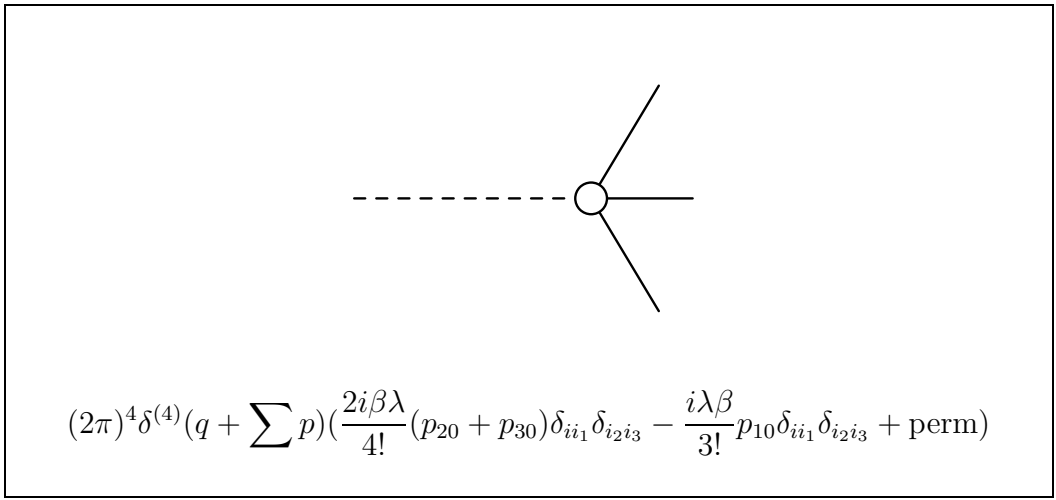

Let us consider the generator 
\be
F_{(\alpha)}=\frac{\lambda \beta}{4!}J^2+\sum_{i=1}^2\alpha_i(\frac{\Pi_i^2}{2}+\frac{m_i^2}{2}\varphi_i^2+
\frac{\lambda}{4!}\varphi_i^2{\vec{\varphi}}^2).\label{w21}
\ee
We decompose $F_{(\alpha )}$, 
\be
F_{(\alpha )}\equiv F_{(\alpha )}^{(0)}+F_{(\alpha )}^{(1)} \label{w22}
\ee
into the $\lambda $-independent and $\lambda$-linear parts:
\be
F_{(\alpha )}^{(0)}&=&\sum_{i=1}^2\alpha_i(\frac{\Pi_i^2}{2}+\frac{m_i^2}{2}\varphi_i^2) \label{w23} \\
F_{(\alpha )}^{(1)}&=&\frac{\lambda}{4!}(\beta J^2+ (\sum_{i=1}^2\alpha_i\varphi_i^2){\vec{\varphi}}^2)
\equiv{}\nonumber\\
&\equiv& \frac{\lambda}{4!}(\beta ({\vec{\varphi}}^2{\vec{\Pi}}^2-(\vec{\varphi}\vec{\Pi})^2)+(\sum_{i=1}^2
\alpha_i\varphi_i^2){\vec{\varphi}}^2). \nonumber
\ee
The same applies to $Q_{(\alpha )i}\equiv \partial F_{(\alpha )}/\partial \Pi_i$:
\be
Q_{(\alpha )i}=Q_{(\alpha )i}^{(0)}+Q_{(\alpha )i}^{(1)} \label{w24}
\ee
with
\be
&&Q_{(\alpha )i}^{(0)}=\alpha_i\Pi_i \label{w25} \\
&&Q_{(\alpha )i}^{(1)}=\frac{-2\lambda \beta}{4!}J\sum_{j=1}^2\varepsilon_{ij}\varphi_j=-\frac{2\lambda \beta}{4!}
((\vec{\Pi} \cdot  \vec{\varphi})\varphi_i-{\vec{\varphi}}^2\Pi_i).\nonumber
\ee
The momentum-space counterpart of eq.(\ref{w20}) reads, in obvious notation,
\be
(-q)F_{i_1...i_n}(q;p_1,...,p_n)=\sum_{k=1}^nQ_{i_k;i_1...\check{i_k}..i_n}(p_k+q;
p_1,..,\check{p_k},...,p_n)\label{w26}
\ee
\begin{fmffile}{obrazki}
We shall give the diagrammatical proof of eq.~(\ref{w26}). The standard Feynman rules for $\phi^4$
theory are shown on fig.~\ref{rul0}.
Further one needs the additional vertices related to the symmetry transformations. They are 
shown on fig.~\ref{rul1}-\ref{rul4} where the notation introduced in eqs.(\ref{w13}) and (\ref{w24}) has been used.
All momenta are directed inwards.

Eq.(\ref{w26}) is represented graphically as follows:
\begin{eqnarray}
\parbox{5cm}{
\fmfframe(5,5)(5,5){
\begin{fmfgraph*}(40,25)
\fmfleft{i1} \fmfrightn{o}{7}
\fmf{dbl_plain,tension=6,width=1.2thick}{v1,v2}
\fmf{dashes}{i1,v1,v2}
\fmflabel{$q$}{i1}
\fmf{phantom,tension=6}{v2,v3}
\fmf{plain,tension=0.5}{v4,o1}
\fmf{phantom,tension=0}{v4,o2}
\fmf{phantom,tension=0}{v4,o3}
\fmf{phantom,tension=0}{v4,o4}
\fmf{phantom,tension=0}{v4,o5}
\fmf{phantom,tension=0}{v4,o6}
\fmf{plain,tension=0.5}{v4,o7}
\fmf{phantom,tension=2}{v3,v4}
\fmflabel{$p_1,i_1$}{o7}
\fmflabel{$p_n,i_n$}{o1}
\fmfv{decor.shape=circ,decor.filled=full,decor.size=0.5thick}{o1,o2,o3,o4,o5,o6,o7}
\fmfv{decor.shape=circ,decor.filled=full,decor.size=5thick}{v3}
\fmfv{decor.shape=circ,decor.filled=empty,decor.size=18thick}{v4}
\end{fmfgraph*}}
}
&=& \sum_k
\parbox{5cm}{
\fmfframe(5,5)(5,5){
\begin{fmfgraph*}(30,25)
\fmfrightn{o}{7}
\fmfleft{i1}
\fmfv{decor.shape=circ,decor.filled=empty,decor.size=5thick}{v1}
\fmf{phantom,tension=6}{i1,v2}
\fmf{phantom,tension=1.5}{v2,v1}
\fmf{plain,tension=0.5}{v2,o1}
\fmf{phantom,tension=0}{v2,o2}
\fmf{phantom,tension=0}{v2,o3}
\fmf{dashes,tension=0.75}{v1,o4}
\fmf{phantom,tension=0}{v2,o5}
\fmf{phantom,tension=0}{v2,o6}
\fmf{plain,tension=0.5}{v2,o7}
\fmflabel{$p_1,i_1$}{o7}
\fmflabel{$p_k+q$}{o4}
\fmflabel{$p_n,i_n$}{o1}
\fmfv{decor.shape=circ,decor.filled=full,decor.size=0.5thick}{o1,o2,o3,o5,o6,o7}
\fmfv{decor.shape=circ,decor.filled=empty,decor.size=22thick}{v2}
\end{fmfgraph*}}
}
\label{w27}
\end{eqnarray}
the blobs stand here for the sums of all relevant Feynman graphs.

We shall give a diagrammatical proof of Ward identities~(\ref{w27}) for arbitrary number
of loops using the Feynman rules described above
(with obvious modifications: $(2\pi)^4\delta^{(4)}(p)\longrightarrow(2\pi)\delta(p)$,
$\int \frac{d^4p}{(2\pi)^4}\ldots\longrightarrow\int\frac{dp}{(2\pi)}$, etc.)

We start from lowest order identities which follow from the invariance of the hamiltonian
to the orders $0,1$ and $2$ in $\lambda$. First, free field theory is invariant under
the canonical transformations generated by $F^{(0)}_{(\alpha)}$. For two-point
Green function we obtain:
\begin{eqnarray}
\parbox{4cm}{
\fmfframe(5,5)(5,5){
\begin{fmfgraph*}(30,25)
\fmfleft{i1} \fmfright{o1,o2}
\fmf{dbl_plain,tension=4,width=1.2thick}{v1,v2}
\fmf{dashes}{i1,v1,v2}
\fmf{phantom,tension=6}{v2,v3}
\fmf{plain,tension=0.5}{v3,o1}
\fmf{plain,tension=0.5}{v3,o2}
\fmflabel{$q$}{i1}
\fmflabel{$p_1,i_1$}{o2}
\fmflabel{$p_2,i_2$}{o1}
\fmfv{decor.shape=circ,decor.filled=full,decor.size=1thick}{o1,o2}
\fmfv{decor.shape=circ,decor.filled=full,decor.size=5thick}{v3}
\end{fmfgraph*}
}}
&=&\kern-40pt
\parbox{4cm}{
\fmfframe(5,5)(5,5){
\begin{fmfgraph*}(30,25)
\fmfleft{i1} \fmfright{o1,o2}
\fmf{phantom}{i1,v1}
\fmf{phantom,tension=2.3}{v1,o1}
\fmf{phantom}{i1,v2}
\fmf{phantom,tension=2.3}{v2,o2}
\fmf{dashes,tension=0}{v2,o2}
\fmf{plain,tension=0,left=0.5}{v1,v2}
\fmflabel{$p_1+q,i_1$}{o2}
\fmfv{decor.shape=circ,decor.filled=empty,decor.size=5thick}{v2}
\fmfv{decor.shape=circ,decor.filled=full,decor.size=1thick}{v1}
\fmfv{label=$p_2\mcom i_2$,label.angle=0}{v1}
\end{fmfgraph*}}}
+\kern-40pt
\parbox{4cm}{
\fmfframe(5,5)(5,5){
\begin{fmfgraph*}(30,25)
\fmfleft{i1} \fmfright{o1,o2}
\fmf{phantom}{i1,v1}
\fmf{phantom,tension=2.3}{v1,o1}
\fmf{phantom}{i1,v2}
\fmf{phantom,tension=2.3}{v2,o2}
\fmf{dashes,tension=0}{v1,o1}
\fmf{plain,tension=0,left=0.5}{v1,v2}
\fmflabel{$p_2+q,i_2$}{o1}
\fmfv{decor.shape=circ,decor.filled=empty,decor.size=5thick}{v1}
\fmfv{decor.shape=circ,decor.filled=full,decor.size=1thick}{v2}
\fmfv{label=$p_1\mcom i_1$,label.angle=0}{v2}
\end{fmfgraph*}}}
\label{w28}
\end{eqnarray}

\vspace{2mm}
This is the only nontrivial identity to this order because all other Green functions are
disconnected. 

To the first order in $\lambda$ we must consider four--point functions. We modify the 
four--vertex by replacing one leg by $Q^{(0)}_{(\alpha)i}$:
\begin{eqnarray}
\parbox{4cm}{
\fmfframe(5,5)(5,5){
\begin{fmfgraph}(30,25)
\fmfleft{i1,i2} \fmfright{o1,o2}
\fmf{plain}{i1,o2}
\fmf{plain}{i2,o1}
\end{fmfgraph}}}
&\longrightarrow&
\parbox{4cm}{
\fmfframe(5,5)(5,5){
\begin{fmfgraph}(30,25)
\fmfleft{i1,i2} \fmfright{o1,o2}
\fmf{plain}{i2,v1,o1}
\fmf{phantom}{i1,v1,o2}
\fmffreeze
\fmf{plain}{v1,o2}
\fmf{plain,tension=0.5}{i1,v2}
\fmf{dashes,tension=0.5}{v2,v1}
\fmfv{decor.shape=circ,decor.filled=empty,decor.size=5thick}{v2}
\end{fmfgraph}}}
\label{w29}
\end{eqnarray}
Then one checks easily that
\begin{equation}
\parbox{4cm}{
\fmfframe(5,5)(5,5){
\begin{fmfgraph}(30,25)
\fmfleft{i1} \fmfright{o1,o2,o3,o4}
\fmf{dbl_plain,tension=4,width=1.2thick}{v1,v2}
\fmf{dashes}{i1,v1,v2}
\fmf{phantom,tension=6}{v2,v3}
\fmf{plain,tension=0.25}{v3,o1}
\fmf{plain,tension=0.25}{v3,o2}
\fmf{plain,tension=0.25}{v3,o3}
\fmf{plain,tension=0.25}{v3,o4}
\fmfv{decor.shape=circ,decor.filled=full,decor.size=1thick}{o1,o2,o3,o4}
\fmfv{decor.shape=circ,decor.filled=full,decor.size=5thick}{v3}
\end{fmfgraph}}}
+\sum\kern-15pt
\parbox{4cm}{
\fmfframe(5,5)(5,5){
\begin{fmfgraph*}(30,25)
\fmfleft{i1,i2} \fmfright{o1,o2}
\fmf{plain}{i1,v1,o2}
\fmf{phantom}{i2,v1,o1}
\fmffreeze
\fmf{plain}{v1,o1}
\fmf{plain}{i2,v2}
\fmf{dashes}{v2,v1}
\fmfv{decor.shape=circ,decor.filled=full,decor.size=1thick}{i1,i2,o1,o2}
\fmfv{decor.shape=circ,decor.filled=full,decor.size=2thick}{v1}
\fmfv{decor.shape=circ,decor.filled=empty,decor.size=5thick}{v2}
\fmfv{label=$p+q$,label.angle=80}{i2}
\end{fmfgraph*}}}
\kern-10pt=\sum\kern-20pt
\parbox{4cm}{
\fmfframe(5,5)(5,5){
\begin{fmfgraph*}(30,25)
\fmfleft{i1,i2} \fmfright{o1,o2}
\fmf{plain}{i1,v1,o2}
\fmf{phantom}{i2,v1,o1}
\fmffreeze
\fmf{plain}{v1,o1}
\fmf{dashes}{i2,v1}
\fmfv{decor.shape=circ,decor.filled=full,decor.size=1thick}{i1,o1,o2}
\fmfv{decor.shape=circ,decor.filled=empty,decor.size=5thick}{v1}
\fmfv{label=$p+q$,label.angle=80}{i2}
\end{fmfgraph*}}}
\label{w30}
\end{equation}
indeed, eq.(\ref{w30}) is simply a statement that all terms of the first order in 
$\lambda$ cancel against each other when eqs.(\ref{w6}) and (\ref{w7}) are inserted
into the hamiltonian~(\ref{w2}).

Eqs. (\ref{w28}) and (\ref{w30}) imply the lowest order counterpart of eq.~(\ref{w27}) for
$n=4$:
\begin{eqnarray}
\kern-20pt
\parbox{4cm}{\fmfframe(5,5)(5,5){
\begin{fmfgraph}(30,25)
\fmfleft{i1} \fmfright{o1,o2,o3,o4}
\fmf{dbl_plain,tension=5,width=1.2thick}{v1,v2}
\fmf{dashes}{i1,v1,v2}
\fmf{phantom,tension=6}{v2,v3}
\fmf{plain,tension=0.25}{v3,o1}
\fmf{plain,tension=0.25}{v3,o2}
\fmf{plain,tension=0.25}{v3,o3}
\fmf{plain,tension=0.25}{v3,o4}
\fmfv{decor.shape=circ,decor.filled=full,decor.size=1thick}{o1,o2,o3,o4}
\fmfv{decor.shape=circ,decor.filled=full,decor.size=5thick}{v3}
\end{fmfgraph}}}
\kern-10pt+\sum\kern-20pt
\parbox{4cm}{
\fmfframe(5,5)(5,5){
\begin{fmfgraph*}(30,25)
\fmfleft{i1,i2} \fmfright{o1,o2}
\fmf{plain}{i2,v1,o1}
\fmf{phantom}{i1,v1,o2}
\fmffreeze
\fmf{plain}{i1,v1}
\fmf{plain}{v1,v2,o2}
\fmfv{decor.shape=circ,decor.filled=full,decor.size=1thick}{i1,i2,o1,o2}
\fmffreeze
\fmftop{i3}
\fmf{dbl_plain,tension=3,width=1.2thick}{v4,v3}
\fmf{phantom,tension=3}{v3,v2}
\fmf{dashes}{i3,v4,v3}
\fmfv{decor.shape=circ,decor.filled=full,decor.size=5thick}{v2}
\end{fmfgraph*}}}
\kern-10pt&=&\sum\kern-20pt
\parbox{4cm}{
\fmfframe(5,5)(5,5){
\begin{fmfgraph*}(30,25)
\fmfleft{i1,i2} \fmfright{o1,o2}
\fmf{plain}{i2,v1,o1}
\fmf{phantom}{i1,v1,o2}
\fmffreeze
\fmf{plain}{i1,v1}
\fmf{plain}{v1,v2}
\fmf{dashes}{v2,o2}
\fmfv{decor.shape=circ,decor.filled=full,decor.size=1thick}{i1,i2,o1}
\fmfv{decor.shape=circ,decor.filled=empty,decor.size=5thick}{v2}
\end{fmfgraph*}}}\nonumber\\
&&{}+\sum\kern-20pt
\parbox{4cm}{
\fmfframe(5,5)(5,5){
\begin{fmfgraph*}(30,25)
\fmfleft{i1,i2} \fmfright{o1,o2}
\fmf{phantom}{i2,v1,o1}
\fmf{phantom}{i1,v1,o2}
\fmffreeze
\fmf{plain}{i1,v1}
\fmf{dashes}{v1,o2}
\fmf{plain}{i2,v2}
\fmf{plain,tension=2,right=0.3}{v2,v1}
\fmf{plain}{o1,v3}
\fmf{plain,tension=2,left=0.3}{v3,v1}
\fmf{phantom}{v2,v3}
\fmf{phantom,tension=0.7}{i1,v2}
\fmf{phantom,tension=0.7}{i1,v3}
\fmfv{decor.shape=circ,decor.filled=full,decor.size=1thick}{i1,i2,o1}
\fmfv{decor.shape=circ,decor.filled=empty,decor.size=5thick}{v1}
\end{fmfgraph*}}}
\label{w31}
\end{eqnarray}

Finally there is an identity due to the cancellation of $\lambda^2$--terms. To write it out 
we modify the four vertex by replacing one leg by
$Q^{(1)}_{(\alpha)i}$:
\begin{equation}
\parbox{4cm}{
\fmfframe(5,5)(5,5){
\begin{fmfgraph*}(30,25)
\fmfleft{i1,i2} \fmfright{o1,o2}
\fmf{plain}{i1,o2}
\fmf{plain}{i2,o1}
\end{fmfgraph*}}}
\longrightarrow
\parbox{4cm}{
\fmfframe(5,5)(5,5){
\begin{fmfgraph*}(30,25)
\fmfleft{i1,i2} \fmfright{o1,o2}
\fmf{plain}{i1,v1,o2}
\fmf{phantom}{i2,v1,o1}
\fmffreeze
\fmf{plain,tension=1}{v1,o1}
\fmf{dashes,tension=2}{v2,v1}
\fmf{plain}{i2,v2}
\fmfv{decor.shape=circ,decor.filled=empty,decor.size=5thick}{v2}
\fmffreeze
\fmf{plain,tension=0}{v3,v2}
\fmf{plain,tension=0}{v4,v2}
\fmf{phantom,tension=1}{v3,i2}
\fmf{phantom,tension=1}{v4,i2}
\fmf{phantom,tension=0.2}{v3,i1}
\fmf{phantom,tension=0.2}{v4,o2}
\end{fmfgraph*}}}
\label{w32}
\end{equation}
The $\lambda^2$ identity reads then:
\begin{equation}
\sum\kern-20pt
\parbox{4cm}{
\fmfframe(5,5)(5,5){
\begin{fmfgraph*}(30,25)
\fmfleft{i1,i2} \fmfright{o1,o2}
\fmf{plain}{i1,v1,o2}
\fmf{phantom}{i2,v1,o1}
\fmffreeze
\fmf{plain,tension=1}{v1,o1}
\fmf{dashes,tension=2}{v2,v1}
\fmf{plain}{i2,v2}
\fmfv{decor.shape=circ,decor.filled=empty,decor.size=5thick}{v2}
\fmffreeze
\fmf{plain,tension=0}{v3,v2}
\fmf{plain,tension=0}{v4,v2}
\fmf{phantom,tension=1}{v3,i2}
\fmf{phantom,tension=1}{v4,i2}
\fmf{phantom,tension=0.2}{v3,i1}
\fmf{phantom,tension=0.2}{v4,o2}
\end{fmfgraph*}}}\kern-10pt=\ 0
\label{w33}
\end{equation}
where the summation goes over all permutations of external lines leading to new graphs.

Now one can give the proof of general identities~(\ref{w26}). Let
\begin{equation}
\parbox{4cm}{
\fmfframe(5,5)(5,5){
\begin{fmfgraph*}(30,25)
\fmfleft{i1} \fmfrightn{o}{7}
\fmf{phantom,tension=2}{i1,v1}
\fmf{plain,tension=0.5}{v1,o1}
\fmf{phantom,tension=0}{v1,o2}
\fmf{phantom,tension=0}{v1,o3}
\fmf{phantom,tension=0}{v1,o4}
\fmf{phantom,tension=0}{v1,o5}
\fmf{phantom,tension=0}{v1,o6}
\fmf{plain,tension=0.5}{v1,o7}
\fmfv{decor.shape=circ,decor.filled=empty,decor.size=24thick}{v1}
\fmfv{decor.shape=circ,decor.filled=full,decor.size=1thick}{o1,o2,o3,o4,o5,o6,o7}
\fmflabel{$p_1,i_1$}{o7}
\fmflabel{$p_n,i_n$}{o1}
\end{fmfgraph*}}}
\label{w34}
\end{equation}
stand for the sum of all graphs of a given order contributing to $n$--point Green function.
The relevant graphs contributing to the left hand side of eq.~(\ref{w26}) to the same
order are obtained either by inserting $F^{(0)}_{(\alpha)}$ into arbitrary line of 
any graph contained in~(\ref{w34}) or by replacing any vertex by $F^{(1)}_{(\alpha)}$.
Therefore one can write
\begin{eqnarray}
\parbox{5cm}{
\fmfframe(5,5)(5,5){
\begin{fmfgraph*}(40,25)
\fmfleft{i1} \fmfrightn{o}{7}
\fmf{dbl_plain,tension=6,width=1.2thick}{v1,v2}
\fmf{dashes}{i1,v1,v2}
\fmflabel{$q$}{i1}
\fmf{phantom,tension=6}{v2,v3}
\fmf{plain,tension=0.5}{v4,o1}
\fmf{phantom,tension=0}{v4,o2}
\fmf{phantom,tension=0}{v4,o3}
\fmf{phantom,tension=0}{v4,o4}
\fmf{phantom,tension=0}{v4,o5}
\fmf{phantom,tension=0}{v4,o6}
\fmf{plain,tension=0.5}{v4,o7}
\fmf{phantom,tension=2}{v3,v4}
\fmflabel{$p_1,i_1$}{o7}
\fmflabel{$p_n,i_n$}{o1}
\fmfv{decor.shape=circ,decor.filled=full,decor.size=1thick}{o1,o2,o3,o4,o5,o6,o7}
\fmfv{decor.shape=circ,decor.filled=full,decor.size=5thick}{v3}
\fmfv{decor.shape=circ,decor.filled=empty,decor.size=18thick}{v4}
\end{fmfgraph*}}}
&=&
\parbox{5.5cm}{
\fmfframe(5,10)(5,10){
\begin{fmfgraph*}(45,25)
\fmfleft{i1} \fmfrightn{o}{7}
\fmf{dbl_plain,tension=6,width=1.2thick}{v1,v2}
\fmf{dashes}{i1,v1,v2}
\fmflabel{$q$}{i1}
\fmf{phantom,tension=6}{v2,v3}
\fmf{plain,tension=0.5}{v4,o1}
\fmf{phantom,tension=0}{v4,o2}
\fmf{phantom,tension=0}{v4,o3}
\fmf{phantom,tension=0}{v4,o4}
\fmf{phantom,tension=0}{v4,o5}
\fmf{phantom,tension=0}{v4,o6}
\fmf{plain,tension=0.5}{v4,o7}
\fmf{phantom,tension=1}{v3,v4}
\fmf{plain,tension=0,left=0.5}{v3,v4}
\fmf{plain,tension=0,right=0.5}{v3,v4}
\fmflabel{$p_1,i_1$}{o7}
\fmflabel{$p_n,i_n$}{o1}
\fmfv{decor.shape=circ,decor.filled=full,decor.size=1thick}{o1,o2,o3,o4,o5,o6,o7}
\fmfv{decor.shape=circ,decor.filled=full,decor.size=5thick}{v3}
\fmfv{decor.shape=circ,decor.filled=empty,decor.size=18thick}{v4}
\end{fmfgraph*}}}+{}\nonumber\\
&+&
\parbox{5.5cm}{
\fmfframe(5,5)(5,5){
\begin{fmfgraph*}(45,25)
\fmfleft{i1} \fmfrightn{o}{7}
\fmf{dbl_plain,tension=6,width=1.2thick}{v1,v2}
\fmf{dashes}{i1,v1,v2}
\fmflabel{$q$}{i1}
\fmf{phantom,tension=6}{v2,v3}
\fmf{plain,tension=0.5}{v4,o1}
\fmf{phantom,tension=0}{v4,o2}
\fmf{phantom,tension=0}{v4,o3}
\fmf{phantom,tension=0}{v4,o4}
\fmf{phantom,tension=0}{v4,o5}
\fmf{phantom,tension=0}{v4,o6}
\fmf{plain,tension=0.5}{v4,o7}
\fmf{phantom,tension=1}{v3,v4}
\fmf{plain,tension=0,left=0.7}{v3,v4}
\fmf{plain,tension=0,left=0.3}{v3,v4}
\fmf{plain,tension=0,right=0.7}{v3,v4}
\fmf{plain,tension=0,right=0.3}{v3,v4}
\fmflabel{$p_1,i_1$}{o7}
\fmflabel{$p_n,i_n$}{o1}
\fmfv{decor.shape=circ,decor.filled=full,decor.size=1thick}{o1,o2,o3,o4,o5,o6,o7}
\fmfv{decor.shape=circ,decor.filled=full,decor.size=5thick}{v3}
\fmfv{decor.shape=circ,decor.filled=empty,decor.size=18thick}{v4}
\end{fmfgraph*}}}
\label{w35}
\end{eqnarray}

Let us consider the first term on the right hand side. Using eq.~(\ref{w28}) one obtains
\begin{eqnarray}
{}&{}&\parbox{5.5cm}{
\fmfframe(5,10)(5,10){
\begin{fmfgraph*}(45,25)
\fmfleft{i1} \fmfrightn{o}{7}
\fmf{dbl_plain,tension=6,width=1.2thick}{v1,v2}
\fmf{dashes}{i1,v1,v2}
\fmflabel{$q$}{i1}
\fmf{phantom,tension=6}{v2,v3}
\fmf{plain,tension=0.5}{v4,o1}
\fmf{phantom,tension=0}{v4,o2}
\fmf{phantom,tension=0}{v4,o3}
\fmf{phantom,tension=0}{v4,o4}
\fmf{phantom,tension=0}{v4,o5}
\fmf{phantom,tension=0}{v4,o6}
\fmf{plain,tension=0.5}{v4,o7}
\fmf{phantom,tension=1}{v3,v4}
\fmf{plain,tension=0,left=0.5}{v3,v4}
\fmf{plain,tension=0,right=0.5}{v3,v4}
\fmflabel{$p_1,i_1$}{o7}
\fmflabel{$p_n,i_n$}{o1}
\fmfv{decor.shape=circ,decor.filled=full,decor.size=1thick}{o1,o2,o3,o4,o5,o6,o7}
\fmfv{decor.shape=circ,decor.filled=full,decor.size=5thick}{v3}
\fmfv{decor.shape=circ,decor.filled=empty,decor.size=18thick}{v4}
\end{fmfgraph*}}}
=\kern-20pt
\parbox{5.5cm}{
\fmfframe(5,10)(5,10){
\begin{fmfgraph*}(45,25)
\fmfleft{i1} \fmfrightn{o}{7}
\fmf{phantom,tension=3}{i1,v1}
\fmf{phantom,tension=1}{v1,v2}
\fmf{plain,tension=0.5}{v2,o1}
\fmf{phantom,tension=0}{v2,o2}
\fmf{phantom,tension=0}{v2,o3}
\fmf{phantom,tension=0}{v2,o4}
\fmf{phantom,tension=0}{v2,o5}
\fmf{phantom,tension=0}{v2,o6}
\fmf{plain,tension=0.5}{v2,o7}
\fmf{dashes,tension=0,left=0.5}{v1,v2}
\fmf{plain,tension=0,right=0.5}{v1,v2}
\fmflabel{$p_1,i_1$}{o7}
\fmflabel{$p_n,i_n$}{o1}
\fmfv{decor.shape=circ,decor.filled=full,decor.size=1thick}{o1,o2,o3,o4,o5,o6,o7}
\fmfv{decor.shape=circ,decor.filled=empty,decor.size=5thick}{v1}
\fmfv{decor.shape=circ,decor.filled=empty,decor.size=18thick}{v2}
\end{fmfgraph*}}}+{}\nonumber\\
&&{}+\kern-20pt
\parbox{5.5cm}{
\fmfframe(5,10)(5,10){
\begin{fmfgraph*}(45,25)
\fmfleft{i1} \fmfrightn{o}{7}
\fmf{phantom,tension=3}{i1,v1}
\fmf{phantom,tension=1}{v1,v2}
\fmf{plain,tension=0.5}{v2,o1}
\fmf{phantom,tension=0}{v2,o2}
\fmf{phantom,tension=0}{v2,o3}
\fmf{phantom,tension=0}{v2,o4}
\fmf{phantom,tension=0}{v2,o5}
\fmf{phantom,tension=0}{v2,o6}
\fmf{plain,tension=0.5}{v2,o7}
\fmf{plain,tension=0,left=0.5}{v1,v2}
\fmf{dashes,tension=0,right=0.5}{v1,v2}
\fmflabel{$p_1,i_1$}{o7}
\fmflabel{$p_n,i_n$}{o1}
\fmfv{decor.shape=circ,decor.filled=full,decor.size=1thick}{o1,o2,o3,o4,o5,o6,o7}
\fmfv{decor.shape=circ,decor.filled=empty,decor.size=5thick}{v1}
\fmfv{decor.shape=circ,decor.filled=empty,decor.size=18thick}{v2}
\end{fmfgraph*}}}
+\ {\displaystyle\sum_k}
\parbox{4.5cm}{
\fmfframe(5,10)(5,10){
\begin{fmfgraph*}(35,25)
\fmfleft{i1} \fmfrightn{o}{7}
\fmf{phantom,tension=4}{i1,v1}
\fmf{plain,tension=0.5}{v1,o1}
\fmf{phantom,tension=0}{v1,o2}
\fmf{phantom,tension=0}{v1,o3}
\fmf{phantom,tension=0}{v1,o4}
\fmf{phantom,tension=0}{v1,o5}
\fmf{phantom,tension=0}{v1,o6}
\fmf{plain,tension=0.5}{v1,o7}
\fmflabel{$p_1,i_1$}{o7}
\fmflabel{$p_n,i_n$}{o1}
\fmfv{decor.shape=circ,decor.filled=full,decor.size=1thick}{o1,o2,o3,o5,o6,o7}
\fmfv{decor.shape=circ,decor.filled=empty,decor.size=18thick}{v1}
\fmffreeze
\fmf{plain,tension=1}{v1,v2}
\fmf{dashes,tension=1.2}{v2,o4}
\fmfv{decor.shape=circ,decor.filled=empty,decor.size=5thick}{v2}
\end{fmfgraph*}}}
\label{w36}
\end{eqnarray}

The last term on the right hand side of eq.~(\ref{w36}) contributes to the right hand side 
of eq.~(\ref{w26}). On the other hand the first two terms together with\nopagebreak{ }
\nopagebreak{}the\nopagebreak{} last term of
\nopagebreak{}q.~(\ref{w35}) can be written as\nopagebreak
\nopagebreak\begin{eqnarray}
\lefteqn{
\parbox{5.5cm}{
\fmfframe(5,5)(5,5){
\begin{fmfgraph}(45,25)
\fmfleft{i1} \fmfrightn{o}{7}
\fmf{phantom,tension=3}{i1,v1}
\fmf{phantom,tension=1}{v1,v2}
\fmf{plain,tension=0.5}{v2,o1}
\fmf{phantom,tension=0}{v2,o2}
\fmf{phantom,tension=0}{v2,o3}
\fmf{phantom,tension=0}{v2,o4}
\fmf{phantom,tension=0}{v2,o5}
\fmf{phantom,tension=0}{v2,o6}
\fmf{plain,tension=0.5}{v2,o7}
\fmf{dashes,tension=0,left=0.5}{v1,v2}
\fmf{plain,tension=0,right=0.5}{v1,v2}
\fmfv{decor.shape=circ,decor.filled=full,decor.size=1thick}{o1,o2,o3,o4,o5,o6,o7}
\fmfv{decor.shape=circ,decor.filled=empty,decor.size=5thick}{v1}
\fmfv{decor.shape=circ,decor.filled=empty,decor.size=18thick}{v2}
\end{fmfgraph}}}
+\kern-20pt
\parbox{5.5cm}{
\fmfframe(5,5)(5,5){
\begin{fmfgraph}(45,25)
\fmfleft{i1} \fmfrightn{o}{7}
\fmf{phantom,tension=3}{i1,v1}
\fmf{phantom,tension=1}{v1,v2}
\fmf{plain,tension=0.5}{v2,o1}
\fmf{phantom,tension=0}{v2,o2}
\fmf{phantom,tension=0}{v2,o3}
\fmf{phantom,tension=0}{v2,o4}
\fmf{phantom,tension=0}{v2,o5}
\fmf{phantom,tension=0}{v2,o6}
\fmf{plain,tension=0.5}{v2,o7}
\fmf{plain,tension=0,left=0.5}{v1,v2}
\fmf{dashes,tension=0,right=0.5}{v1,v2}
\fmfv{decor.shape=circ,decor.filled=full,decor.size=1thick}{o1,o2,o3,o4,o5,o6,o7}
\fmfv{decor.shape=circ,decor.filled=empty,decor.size=5thick}{v1}
\fmfv{decor.shape=circ,decor.filled=empty,decor.size=18thick}{v2}
\end{fmfgraph}}}+{}}&&\nonumber\\
\lefteqn{{}+\kern-10pt
\parbox{5.5cm}{
\fmfframe(5,5)(5,5){
\begin{fmfgraph}(45,25)
\fmfleft{i1} \fmfrightn{o}{7}
\fmf{dbl_plain,tension=6,width=1.2thick}{v1,v2}
\fmf{dashes}{i1,v1,v2}
\fmf{phantom,tension=6}{v2,v3}
\fmf{plain,tension=0.5}{v4,o1}
\fmf{phantom,tension=0}{v4,o2}
\fmf{phantom,tension=0}{v4,o3}
\fmf{phantom,tension=0}{v4,o4}
\fmf{phantom,tension=0}{v4,o5}
\fmf{phantom,tension=0}{v4,o6}
\fmf{plain,tension=0.5}{v4,o7}
\fmf{phantom,tension=1}{v3,v4}
\fmf{plain,tension=0,left=0.7}{v3,v4}
\fmf{plain,tension=0,left=0.3}{v3,v4}
\fmf{plain,tension=0,right=0.7}{v3,v4}
\fmf{plain,tension=0,right=0.3}{v3,v4}
\fmfv{decor.shape=circ,decor.filled=full,decor.size=1thick}{o1,o2,o3,o4,o5,o6,o7}
\fmfv{decor.shape=circ,decor.filled=full,decor.size=5thick}{v3}
\fmfv{decor.shape=circ,decor.filled=empty,decor.size=18thick}{v4}
\end{fmfgraph}}}={}}&&\nonumber\\ 
&=&{\displaystyle\sum}\kern-7pt
\parbox{4.5cm}{
\fmfframe(5,5)(5,5){
\begin{fmfgraph}(35,25)
\fmfleft{i2,i1,i3} \fmfrightn{o}{7}
\fmf{phantom,tension=1.5}{i1,v3,v1}
\fmf{plain,tension=0.5}{v1,o1}
\fmf{phantom,tension=0}{v1,o2}
\fmf{phantom,tension=0}{v1,o3}
\fmf{phantom,tension=0}{v1,o4}
\fmf{phantom,tension=0}{v1,o5}
\fmf{phantom,tension=0}{v1,o6}
\fmf{plain,tension=0.5}{v1,o7}
\fmfv{decor.shape=circ,decor.filled=empty,decor.size=18thick}{v1}
\fmfv{decor.shape=circ,decor.filled=full,decor.size=1thick}{o1,o2,o3,o4,o5,o6,o7}
\fmffreeze
\fmf{plain,right=0}{i1,v1}
\fmf{plain,right=0.4}{i1,v1}
\fmf{plain,right=0.85}{i1,v1}
\fmf{phantom,tension=1}{i3,v2}
\fmf{phantom,tension=1}{v2,v3}
\fmffreeze
\fmf{dashes,left=0.3}{i1,v2}
\fmf{plain,left=0.5}{v2,v1}
\fmfv{decor.shape=circ,decor.filled=empty,decor.size=5thick}{v2}
\fmfv{decor.shape=circ,decor.filled=full,decor.size=0.7thick}{i1}
\end{fmfgraph}}}
\kern-5pt + \kern-10pt
\parbox{5.5cm}{
\fmfframe(5,5)(5,5){
\begin{fmfgraph}(45,25)
\fmfleft{i1} \fmfrightn{o}{7}
\fmf{dbl_plain,tension=6,width=1.2thick}{v1,v2}
\fmf{dashes}{i1,v1,v2}
\fmf{phantom,tension=6}{v2,v3}
\fmf{plain,tension=0.5}{v4,o1}
\fmf{phantom,tension=0}{v4,o2}
\fmf{phantom,tension=0}{v4,o3}
\fmf{phantom,tension=0}{v4,o4}
\fmf{phantom,tension=0}{v4,o5}
\fmf{phantom,tension=0}{v4,o6}
\fmf{plain,tension=0.5}{v4,o7}
\fmf{phantom,tension=1}{v3,v4}
\fmf{plain,tension=0,left=0.7}{v3,v4}
\fmf{plain,tension=0,left=0.3}{v3,v4}
\fmf{plain,tension=0,right=0.7}{v3,v4}
\fmf{plain,tension=0,right=0.3}{v3,v4}
\fmfv{decor.shape=circ,decor.filled=full,decor.size=1thick}{o1,o2,o3,o4,o5,o6,o7}
\fmfv{decor.shape=circ,decor.filled=full,decor.size=5thick}{v3}
\fmfv{decor.shape=circ,decor.filled=empty,decor.size=18thick}{v4}
\end{fmfgraph}}}
\label{w37}
\end{eqnarray}

By the identity~(\ref{w30}) the last sum can be rewritten as
\vspace{1.5mm}
\begin{equation}
\sum_k
\parbox{5.5cm}{
\fmfframe(5,5)(15,5){
\begin{fmfgraph*}(35,25)
\fmfleft{i1} \fmfrightn{o}{7}
\fmf{phantom,tension=4}{i1,v1}
\fmf{plain,tension=0.5}{v1,o1}
\fmf{phantom,tension=0}{v1,o2}
\fmf{phantom,tension=0}{v1,o3}
\fmf{phantom,tension=0}{v1,o4}
\fmf{phantom,tension=0}{v1,o5}
\fmf{phantom,tension=0}{v1,o6}
\fmf{plain,tension=0.5}{v1,o7}
\fmflabel{$p_1,i_1$}{o7}
\fmflabel{$p_n,i_n$}{o1}
\fmfv{decor.shape=circ,decor.filled=full,decor.size=1thick}{o1,o2,o3,o5,o6,o7}
\fmfv{decor.shape=circ,decor.filled=empty,decor.size=25thick}{v1}
\fmffreeze
\fmf{phantom,tension=1}{v1,v2}
\fmf{phantom,tension=1.5}{v2,o4}
\fmflabel{$p_k+q$}{o4}
\fmffreeze
\fmf{dashes}{v2,o4}
\fmf{plain,left=0}{v1,v2}
\fmf{plain,left=0.3}{v1,v2}
\fmf{plain,right=0.3}{v1,v2}
\fmfv{decor.shape=circ,decor.filled=empty,decor.size=5thick}{v2}
\end{fmfgraph*}}}
+\sum\kern-10pt
\parbox{5.5cm}{
\fmfframe(5,5)(5,5){
\begin{fmfgraph*}(45,25)
\fmfleft{i2,i1,i3} \fmfrightn{o}{7}
\fmf{phantom,tension=1}{i1,v1}
\fmf{plain,tension=0.5}{v1,o1}
\fmf{phantom,tension=0}{v1,o2}
\fmf{phantom,tension=0}{v1,o3}
\fmf{phantom,tension=0}{v1,o4}
\fmf{phantom,tension=0}{v1,o5}
\fmf{phantom,tension=0}{v1,o6}
\fmf{plain,tension=0.5}{v1,o7}
\fmflabel{$p_1,i_1$}{o7}
\fmflabel{$p_n,i_n$}{o1}
\fmfv{decor.shape=circ,decor.filled=full,decor.size=1thick}{o1,o2,o3,o4,o5,o6,o7}
\fmfv{decor.shape=circ,decor.filled=empty,decor.size=25thick}{v1}
\fmffreeze
\fmf{phantom,tension=2.5}{i2,v2}
\fmf{phantom,tension=1}{v2,v1}
\fmf{phantom,tension=2.5}{i3,v3}
\fmf{phantom,tension=1}{v3,v1}
\fmf{phantom,tension=0.5}{i1,v2}
\fmf{phantom,tension=0.5}{i1,v3}
\fmffreeze
\fmf{dashes,left=0.35}{i1,v3}
\fmf{dashes,right=0.5}{i1,v2}
\fmf{plain,right=0.4}{v2,v1}
\fmf{plain,right=0.55}{v2,v1}
\fmf{plain,right=0.7}{v2,v1}
\fmf{plain,left=0.35}{v3,v1}
\fmf{plain,left=0.55}{v3,v1}
\fmf{plain,left=0.7}{v3,v1}
\fmfv{decor.shape=circ,decor.filled=empty,decor.size=5thick}{v2}
\end{fmfgraph*}}}
\label{w38}
\end{equation}
\vspace{3mm}
The last term vanishes due to the identity~(\ref{w33}). Collecting together 
eqs.~(\ref{w34})--(\ref{w38}) one obtains the basic identity~(\ref{w26}).

\end{fmffile}


\section{Nullification}

The threshold tree-graph Green functions for fourdimensional theory coincide (up to some irrelevant
factors) with those of our reduced theory. Therefore the Ward identities (\ref{w27}) hold for those 
Green functions. Consider the connected tree amplitude with $n_1$\ initial $\varphi_{1}$-
and $n_2$\ final $\varphi_{2}$-lines; here $n_1$\ and $n_2$\ are coprime numbers up to one common
divisor $2$\ such that $n_1m_1=n_2m_2$. Consider the Ward identity (\ref{w27}) with $n_1\;\;
\varphi_1$-lines and $n_2\;\;\varphi_2$-lines, $n=n_1+n_2$. It is easy to see that, due to the definition
of $n_1$\ and $n_2$, the following properties hold:\\
(i) one can put $q=0$\ making the left-hand side of eq.(\ref{w27}) vanishing;\\
(ii) amputating external propagators and passing to mass-shell limit makes the $Q_{(\alpha )i}^{
(1)}$\ contributions vanishing.\\
Taking all that into account we obtain from eq.(\ref{w27})
\be
(\alpha_1n_1m_1-\alpha_2n_2m_2)A(n_1\varphi_1\rightarrow n_2\varphi_2)=0 \label{w37}
\ee
the parameters $\alpha_1, \; \alpha_2$\ being arbitrary. Therefore 
\be
A(n_1\varphi_1\rightarrow n_2\varphi_2)=0 \label{w38}
\ee

The above proof allows for some insight into the cancellation mechanism. It is also obvious 
that such a cancellation works in a wider context: it is sufficient that the reduced system posses a
symmetry containing a linear part in canonical variables which survives the amputation of external
lines propagators. This conclusion has been obtained in the framework of Ref.(\ref{w8}).

\vspace{1cm}
{\noindent\bf\Large Acknowledgement}\linebreak

\noindent I would like to thank 
Prof. Piotr Kosi\'nski for many helpful and fruitful discussions and suggestions.

\end{document}